\documentclass[twocolumn,prb,aps,nobalancelastpage,superscriptaddress,citeautoscript]{revtex4-1}

\usepackage{graphicx}
\usepackage{bm}

\begin{document}

\preprint{v3}
\title{Anisotropic fluctuations and quasiparticle excitations in FeSe$_{0.5}$Te$_{0.5}$}

\author{A. Serafin}
\affiliation{H.H. Wills Physics Laboratory, University of Bristol, Tyndall Avenue, Bristol BS8 1TL, United Kingdom.}
\author{A. I. Coldea}
\altaffiliation{Present Address: Clarendon Laboratory, Department of Physics, University of Oxford, Oxford OX1 3PU,
United Kingdom.}
\affiliation{H.H. Wills Physics Laboratory, University of Bristol, Tyndall Avenue, Bristol BS8 1TL,
United Kingdom.}
\author{A.Y. Ganin}
\affiliation{Department of Chemistry, University of Liverpool, Liverpool L69 7ZD, United Kingdom.}
\author{M.J. Rosseinsky}
\affiliation{Department of Chemistry, University of Liverpool, Liverpool L69 7ZD, United Kingdom.}
\author{K. Prassides}
\affiliation{Department of Chemistry, University Science Laboratories South Road, Durham DH1 3LE, United Kingdom.}
\author{D. Vignolles}
\affiliation{Laboratoire National des Champs Magn\'{e}tiques Intenses (CNRS), 143 Avenue de Rangueil, 31400 Toulouse,
France.}
\author{A. Carrington}
\affiliation{H.H. Wills Physics Laboratory, University of Bristol, Tyndall Avenue, Bristol BS8 1TL, United Kingdom.}

\date{\today}

\begin{abstract}
We  present data for the temperature dependence of the magnetic penetration depth $\lambda(T)$, heat capacity $C(T)$,
resistivity $\rho(T)$ and magnetic torque $\tau$ for highly homogeneous single crystal samples of
Fe$_{1.0}$Se$_{0.44(4)}$Te$_{0.56(4)}$.  $\lambda(T)$ was measured down to 200\,mK in zero field. We find $\lambda(T)$
follows a power law $\Delta\lambda \sim T^n$ with $n=2.2\pm0.1$. This is similar to some 122 iron-arsenides and likely
results from a sign-changing pairing state combined with strong scattering. Magnetic fields of up to $B=$55\,T or 14\,T
were used for the $\tau(B)$ and $C(T)$/$\rho(T)$ measurements respectively. The specific heat, resistivity and torque
measurements were used to map out the $(H,T)$ phase diagram in this material. All three measurements were conducted on
exactly the same single crystal sample so that the different information revealed by these probes is clearly
distinguished. Heat capacity data strongly resemble those found for the high $T_c$ cuprates, where strong fluctuation
effects wipe-out the phase transition at $H_{c2}$. Unusually, here we find the fluctuation effects appear to be
strongly anisotropic.
\end{abstract}
\maketitle

\section{Introduction}

The superconducting iron-chalcogenide compounds  Fe$_{1+y}$Se$_{1-x}$Te$_{x}$ have attracted much interest because of
their many similarities to the high $T_c$ iron-pnictides.  The transition temperature can be varied between  8\,K and
14\,K by changing the chalcogenide ratio $x$ and reaches a maximum at around $x=0.5$ ($y\simeq 0$)
\cite{YehHHCHWLCCLYW08}. These materials share the structural motif of square planar sheets of tetrahedrally
coordinated Fe with the iron-pnictide 1111 and 122 compounds (e.g., SmFeAsO$_{1-x}$F$_x$ with $T_c$=55\,K and
\cite{RenLYYSLCDSZZ08} and  Ba$_{1-x}$K$_x$Fe$_2$As$_2$ with $T_c$ up to 38\,K \cite{RotterTJ08}). The
iron-chalcogenide compounds are structurally simpler than their iron-pnictide counterparts because there are no guest
ions or interleaved layers separating the iron-chalcogenide layers. Like the iron-pnictides the Fermi surface of these
compounds is composed of quasi-two-dimensional electron and hole pockets located at the Brillouin zone corner ($M$
point) and center ($\Gamma$) point respectively \cite{SubediZSD08,TamaiGRBMKCSMPRB10}. It is likely therefore, that if
the unusual Fermi surface topology is the ultimate origin of the pairing interaction then we might expect the mechanism
to be similar in these materials.

In this paper,  we report measurements of two aspects of the physics of the highest $T_c$  member ($x\simeq 0.5$) of
the Fe$_{1+y}$Se$_{1-x}$Te$_{x}$ series.  Firstly, we show data for the magnetic penetration depth $\lambda(T)$ in the
superconducting state.  These measurements give information about the pairing interactions which drive the
superconductivity. Secondly, we show measurements of the heat capacity, magnetic torque and electrical resistivity of
exactly the same single crystal sample.  These measurements are used to derive the temperature-field phase diagram of
this material.  By performing the measurements on the same sample, the different points on the phase diagram measured
by these probes is made clear.  These measurements show several unusual features, most notably, that unlike some other
iron-based superconductors the thermodynamics of the superconducting transition are dominated by strong thermal
fluctuation effects (like the high $T_c$ cuprate superconductors). However, unlike other superconductors these
fluctuations effects appear to be strongly anisotropic.

A powerful way to understand what drives the superconducting pairing interaction is to determine the $\bm{k}$ dependent
structure of the superconducting energy gap $\Delta$. If the pairing interaction $v_{\bm{kk^\prime}}$ is repulsive in
some (or all) directions in $\bm{k}$-space, as expected for most spin-fluctuation driven pairing models, then it is
likely that $\Delta$ will change sign at some point on the Fermi surface \cite{Scalapino95}.  In materials where there
is a single sheet of Fermi surface this inevitably results in gap nodes, i.e., points on the Fermi surface where the
energy gap is zero. However, in the iron-based materials the existence of disconnected electron and hole sheet of Fermi
surface means that the sign-change can be accommodated between two different Fermi surface sheets so that there are
no-nodes on either \cite{MazinSJD08,KurokiOAUTKA08,CvetkovicT09,ChubukovEE08}. Such as state which is known as the
$s_\pm$ state appears to be preferred if the scattering is dominated by wavevectors close to $\bm{q}=(\pi,\pi)$.
However if there is strong intra-band scattering (with low $\bm{q}$) then a state with nodes on the hole and / or
electron sheets may be favored \cite{GraserMHS09,KurokiUOAA09,ChubukovVV09}.

Experiments to determine the gap anisotropy in the iron-based compounds have been conducted on many different
iron-based superconductors.  A variety of different behaviors have been found which might be explained by a combination
of varying levels of sample disorder and in some cases intrinsic differences in the pairing state.  The first single
crystal measurements of the temperature dependent penetration depth were reported for the 1111 family compounds
PrFeAsO$_{1-x}$ ($T_c$=35\,K)\cite{Hashimoto08} and SmFeAsO$_{1-x}$F$_x$ ($T_c$=44\,K) \cite{MaloneFSCZBKK09}. These
measurements showed that for $T\lesssim 0.2 T_c$ $\lambda(T)$ had a very weak temperature dependence compatible with a
fully gapped state. Fitting the full temperature dependence (up to $T_c$) of the superfluid density required a model
with two isotropic ($s$-wave) gaps. Experiments on the Co-doped Ba122 compound, Ba(Fe$_{1-x}$Co$_x$)$_2$As$_2$
($T_c=$13-24\,K) \cite{Gordon08b} showed a robust power-law behavior of $\lambda(T)\propto T^n$ with exponent $n$ in
the range 2--2.5, which might have been explained by a nodal gap in the presence of strong
disorder.\cite{HirschfeldG93} However, measurements \cite{LuoPRB09} of thermal conductivity $\kappa$ did not find a
substantial value of $\kappa/T$ in the $T=0$ limit as would be expected for a sign changing line node in the gap and
theoretical calculations \cite{vorontsov:140507,Bang09,Kogan10,Gordon10} suggested that the observed power-law in
$\lambda(T)$ could result from the influence of strong-pair breaking inter-band scattering on the intrinsically fully
gapped $s_\pm$ state. Strong disorder may be unavoidable in these systems because of the doping of Co into the
conducting FeAs plane. On the other hand, LaFePO is a stoichiometric system in which the quasiparticle mean free paths
are long enough for quantum oscillations to be observable at relatively low field \cite{ColdeaFCABCEFHM08}. In LaFePO,
$\lambda(T)$ was found to vary close to linearly with temperature which is strongly indicative of line nodes in the
clean limit \cite{FletcherPRL09}. Although LaFePO has a relatively low $T_c$ of 6\,K, similar behavior was also seen in
BaFe$_2$(As$_{0.66}$P$_{0.33}$)$_2$ with $T_c$=30\,K \cite{Hashimoto0907.4399}. In both cases a large value of
$\kappa/T|_{T\rightarrow 0}$ was found,\cite{Yamashita09,Hashimoto0907.4399} which further supports the conclusion that
at least some of the Fermi surface sheets have line nodes in these compounds.  So the evidence suggests that the
pairing state is not the same in all iron-based superconductors.  It has been suggested that the different behaviors
are driven by the orbital character of the hole sheets which are in turn determined by the height of the pnictogen atom
\cite{KurokiUOAA09}.  Kogan \cite{Kogan10} has recently shown that if strong pair-breaking is responsible for the $T^2$
power-law behavior of $\lambda(T)$ in the iron-pnictides then the coefficient of this $T^2$ term should be related to
the height of the mean-field part of the specific heat jump at $T_c$ and the slope of the upper critical field
$dH_{c2}/dT$.  We use our data for FeSe$_{0.5}$Te$_{0.5}$ to make a quantitative test of this theory.

\section{Experimental details}

The single crystals used in this work were grown from the elements using iodine as the vapor transport medium at
680$^\circ$C targeting the nominal composition Fe$_{1.0}$Se$_{0.5}$Te$_{0.5}$. The details of the synthesis are
reported elsewhere \cite{Ganin}.  The crystal structure of the exact sample measured here (sample 1) was determined by
single crystal x-ray diffraction. No extra iron was observed between the Fe(Se/Te) slabs so the mean formula was
identified as Fe$_{1.0}$Se$_{0.44(4)}$Te$_{0.56(4)}$. This was additionally confirmed by energy dispersive x-ray (EDX)
analysis.

The magnetic penetration depth as function of temperature has been measured in three different single crystals (typical
 dimensions 300$\times$ 400 $\times$ 4\,$\mu$m$^3$ with the smallest dimension along the $c$-axis), using a
radio frequency tunnel diode oscillator \cite{CarringtonGKG99}. The sample was mounted with vacuum grease on a cold
finger sapphire rod which was inserted into a copper wound solenoid forming the inductor of a resonant $LC$ circuit
with frequency close to 12\,MHz. The RF field was orientated along the $c$-axis of the crystal, thus only in-plane
screening currents were excited. The experiment was shielded from external fields using a mu-metal can and the probe
field was $\lesssim 10^{-6}$\,T  thus minimizing any contributions from mobile vortices.  Two different experimental
set-ups were used. One is mounted in a pumped $^4$He cryostat with base temperature $\sim$1.4\,K and the second in a
dilution refrigerator where the sample may be cooled below 100\,mK (here the resonant frequency is $\sim$14\,MHz)
\cite{Fletcher07}. The calibration factor relating the measured frequency shifts with temperature to $\Delta\lambda(T)$
was determined from the geometry of the sample, and the total perturbation to the resonant frequency due to the sample,
found by withdrawing the sample from the coil at low temperature.  The procedure is described in more detail in Ref.\
\onlinecite{ProzorovGCA00}.

The heat capacity of sample 1 from the penetration depth study (mass $\simeq 3 \mu$g) was measured using a temperature
modulation technique. We use light as a heat source to minimize addenda and make the heating uniform over the sample
surface, however this does mean that the absolute power (and hence heat capacity) are difficult to estimate accurately,
so our data for $C$ are left in arbitrary units. Also the addenda are difficult to determine as the length of
thermocouple wire which contributes depends on the presence or absence of the sample. Despite these limitations the
method is able to measure very small samples  of order 1 microgram over a wide temperature range with very high
resolution. These smaller samples are usually much more homogeneous with sharper superconducting transitions than
larger samples. The sharp transition was critical for the accurate characterization of the fluctuation effects as will
be explained in more detail below. As our present study is focused on studying the evolution of the nature of the
superconducting phase transition with magnetic field the lack of absolute scale for $C$ is not a serious limitation.

The sample was glued with a small amount of VGE-7031 varnish to a flattened thermocouple made from 12$\mu$m diameter
Chromel and Constantan wires (type E) \cite{omega}. The sample and thermocouple couple were then attached to a
temperature controller stage in a vacuum enclosure which sits in the $^4$He bath of a 14\,T superconducting solenoid.
The sample was heated using a modulated light source which was generated by an infra-red light emitting diode at room
temperature and directed onto the sample using a plastic optic fibre.  The size of the temperature oscillations
$T_{ac}$ is inversely related to the heat capacity $C$ of the samples, $T_{ac}=P/\omega C \times f(\omega)$, where $P$
is the power of the sinusoidal heat source (light) at frequency $\omega$ \cite{SullivanS68}.  The function $f(\omega)$
depends on the time constants of the set-up and the operating frequency (typically 6-20\,Hz) is chosen to be the
frequency where $\omega T_{ac}$ has the minimum frequency dependence in the temperature range of interest
\cite{SullivanS68}. This ensures that $f(\omega)$ remains relatively constant as the experimental time constants change
as a function of temperature and therefore $C\propto T_{ac}^{-1}$. The temperature of the LED is stabilized to minimize
drift. The temperature of the stage was monitored using a Cernox 1050 thermometer, the magnetoresistance of which was
corrected for using the data of Ref.\ \onlinecite{BrandtLR99} (although this is negligible in the temperature range of
interest for the present study). In zero magnetic field the thermopower of the thermocouple was taken from the standard
table for type E thermocouples \cite{typeE}.  The field dependence of the thermocouple sensitivity was estimated by
measuring the field dependence of the temperature oscillations when a high purity Ag sample (coated with a thin layer
of black ink to maximize the light absorption) replaces the sample on the same thermocouple.   At $T$=15\,K the
sensitivity changes by $\sim$4\% between zero and 14\,T whereas at $T$=24\,K the change is $\sim$2\%.

Electrical resistivity was measured on a small piece cut from sample 1 (dimensions 310$\times$ 75 $\times$
4\,$\mu$m$^3$)  using the standard four probe method with an ac current excitation of 100$\mu$A. Contacts were made
using pressed Au, which resulted in contact resistances of a few ohms.

Magnetic torque was measured using a piezo resistive cantilever technique \cite{seiko} on another small piece cut from
 sample 1 (dimensions 70$\times$ 100 $\times$ 4\,$\mu$m$^3$).  Here the resistance of an atomic force microscopy
cantilever is measured using Wheatstone bridge arrangement with an ac excitation current of 10$\mu$A at 72\,Hz. The
sample was attached to the cantilever with vacuum grease and mounted in vacuum on a rotating platform in the 14\,T
magnet.  Checks were made at different current levels to ensure that there is no significant self-heating of cantilever
and sample in the temperature range of interest.  Samples from the same batch were also measured using the same
technique but with pulsed fields up to 55\,T in Toulouse. Here the sample/cantilever is surrounded by $^4$He exchange
gas or liquid.

\section{Results}
\subsection{Magnetic penetration depth}

\begin{figure}
\center
\includegraphics[width=8cm]{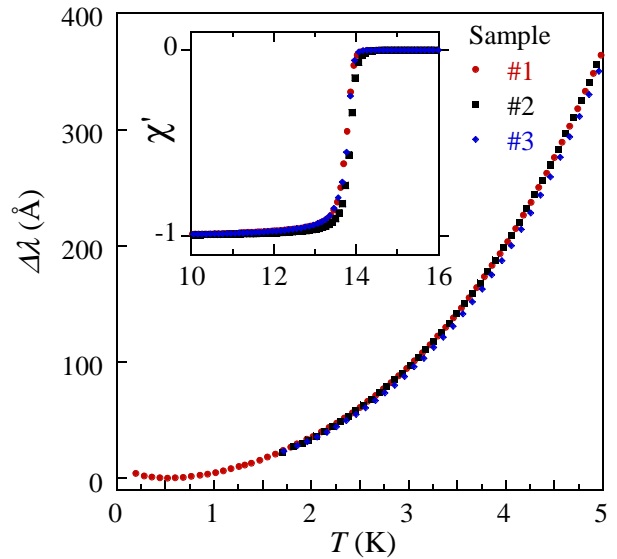}
\caption{(color online) Temperature dependent in-plane penetration depth for three samples of FeSe$_{0.5}$Te$_{0.5}$.
For sample 1 $\Delta \lambda$ is relative to the minimum value measured (at $T$=0.52\,K), whereas for samples 2 and 3
the data have been shifted to coincide with sample 1 at $T$=1.7\,K which was the lowest temperature to which these
samples were measured.  The inset shows the RF susceptibility (normalized to $-1$ at $T\simeq 0$ and 0 in the normal
state) close to $T_c$. The three samples show sharp superconducting transitions ($\Delta T_c\sim 0.5$\,K).}
\label{Fig:RawlamdbaT}
\end{figure}

\begin{figure}
\center
\includegraphics[width=8cm]{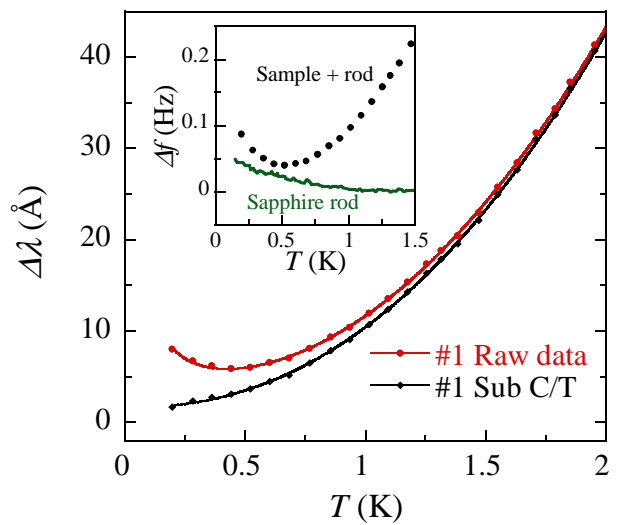}
\caption{(color online) Low temperature behavior of the penetration depth in sample 1 from Fig.\ \ref{Fig:RawlamdbaT}.
The main figure shows the temperature dependence of the sample with the background from the sapphire rod subtracted
(red circles) and also with the fitted paramagnetic term subtracted (black diamonds).  The lines are fits to a power
law (black line) and power law plus Curie term (red line) as described in the text. The inset shows the measured
oscillator frequency shifts for the sample plus sapphire rod and also the sapphire rod alone.} \label{Fig:LambdaLowT}
\end{figure}

Data for the temperature dependence of $\lambda$ for three of our samples of FeSe$_{0.5}$Te$_{0.5}$ are shown in Fig.\
\ref{Fig:RawlamdbaT}.  All three samples show a very sharp superconducting transition at $T_c$=13.8\,K (midpoint)
($\Delta T \sim 0.5$\,K), and very similar low temperature behavior (only sample 1 was measured below 1.7\,K).  The
absolute temperature dependence of the $\Delta \lambda(T)$ is quite reproducible (to within 10\%) indicating that the
calibration factor has been accurately determined and that there are no complications from rough edges which can lead
to an overestimation of the absolute scale of $\Delta\lambda(T)$ \cite{Hashimoto10}.

The low temperature behavior is shown more clearly in Fig.\ \ref{Fig:LambdaLowT}.  Below $T\simeq 0.5$\,K there is a
small upturn in $\Delta\lambda(T)$.  Note that the contribution from the sapphire rod on which the sample is mounted is
smaller and has a significantly weaker temperature dependence (see inset to Fig.\ \ref{Fig:LambdaLowT}).  The
contribution from the rod has been subtracted from the data in the main panel.   The likely origin of this upturn is a
Curie-like paramagnetism of the normal state. In general the measured penetration depth $\lambda_m$ is related to the
London depth ($\lambda_L$) by $\lambda_m(T)=\lambda_L(T)\sqrt{1+\chi_N(T)}$ where $\chi_N(T)$ is the normal state
susceptibility \cite{Cooper96}. Assuming a simple Curie law form for $\chi_N(T)$ then for $\chi_N(T)\ll 1$ the
additional contribution to the measured penetration depth is given by
\begin{equation}
\Delta\lambda_{NP}=\frac{n_i\lambda_L(0)\mu_0\mu_e^2}{6 k_B V_{\rm cell}T}
\end{equation}
where $n_i$ is the number of magnetic ions per unit cell, $V_{\rm cell}$ is the unit cell volume (=86.2\AA$^3$) and
$\mu_e$ is the effective magnetic moment of the paramagnetic ion.   In Fig.\ \ref{Fig:LambdaLowT} we show a fit of the
low temperature $\lambda(T)$ data to a power law dependence plus the $\Delta\lambda_{NP}$ contribution
\begin{equation}
\Delta \lambda(T) = A T^n + \frac{\mathcal{C}}{T} + \lambda_{\rm offset}. \label{Eq:lamTfit}
\end{equation}
This equation fits the data very well, with $n=2.2 \pm 0.1$, $\mathcal{C}=1.24$\AA/K.  This value of $\mathcal{C}$
corresponds to an average moment of 0.16$\mu_B$ per unit cell (assuming $\lambda_L(0)=5000$\,\AA)\cite{BiswasBPTLH10}.
A fit using a Curie-Weiss form $\chi_N=\mathcal{C}/(T+T_\theta)$ gives a slightly better fit (not shown) however as the
fitted value of $T_\theta=0.16$\,K is below our minimum temperature this should be viewed with some caution (for this
fit $\mathcal{C}=3.0$\,\AA/K and the exponent $n$ was unchanged). This small moment could come from the bulk Fe atoms
or possibly from a small amount of free Fe impurities (although we note that none were detected in this sample by the
XRD or EDX measurements).

The observed power law is close to 2, and as discussed above could either indicate a nodal state in the presence of
disorder or instead could result from strong interband scattering between two intrinsically fully gapped Fermi surface
sheets with sign changing gap between the sheets ($s_\pm$ state).  Thermal conductivity measurements
\cite{DongGZQDZPXL09} on the related compound FeSe$_x$ ($T_c=8.8$\,K) show that $\kappa/T|_{T\rightarrow 0}$ is less
than 4\% of its normal-state value suggesting the absence of sign-changing nodes at this composition.  As far as we are
aware there have been no $\kappa(T)$ data published for FeSe$_{0.5}$Te$_{0.5}$, so for this composition the existence
or not of nodes is still an open question.

Previously, penetration depth measurements have been performed by the muon-spin rotation/ relaxation  technique as well
as a tunnel diode oscillator (TDO) technique similar to that used here.  The $\mu$SR results of Bendele \emph{et al.}
\cite{Bendele10} for $x=0.5$ ($T_c$=14\,K) were interpreted using a two gap $s$-wave model, however the data show no
evidence for saturation of $\lambda^{-2}(T)$ at low $T$ due to the lack of data below $T=1.5$\,K ($T/T_c\simeq 0.1$).
The $\mu$SR data of Biswas {\it et al.} \cite{BiswasBPTLH10} for a similar composition did show saturation at low
temperature and were also fitted by a two gap $s$-wave model.  However, the fact that $\mu$SR measurements are
conducted in the mixed state means that these results are not definitive \cite{SonierBKMMBCHHL99,BiswasBPTLH10}. The
TDO measurements of Kim \textit{et al.} \cite{Kim1001.2042} were reported on samples with $x=0.37$ ($T_c\sim 14$\,K).
These data show a power law $\lambda \sim T^n$ down to the lowest temperatures ($T\simeq 0.5$\,K) with $n=2.1$ which is
consistent with our findings for FeSe$_{0.5}$Te$_{0.5}$.

\subsection{Heat Capacity}

\begin{figure}
\center
\includegraphics[width=8cm]{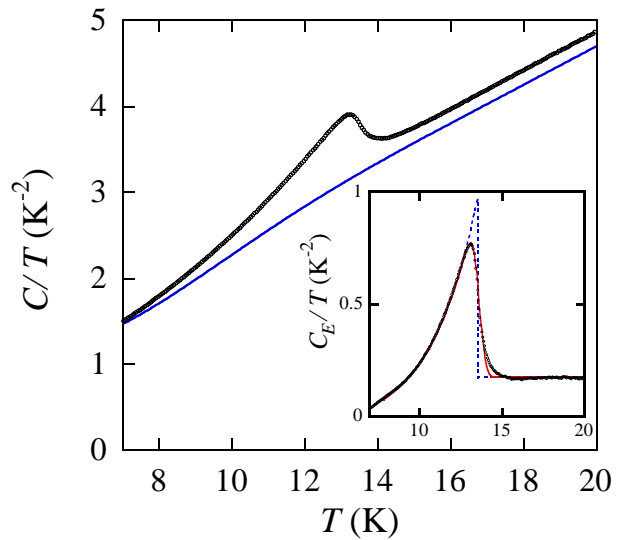}
\caption{(color online) Zero field total heat capacity data. Here we have set the constant of proportionality between
$C$ and $T_{ac}^{-1}$ to unity so the size of the applied temperature oscillations can be read off the axis.  The solid
blue curve is the estimated non-electronic background coming from the sample phonon heat capacity and also the addenda.
The inset shows the electronic part of $C$ with the background subtracted together with the alpha-model fit (blue
curve). The green curve shows the same fit convoluted with a Gaussian $T_c$ distribution, $ T_c^0 = 13.60$\,K and
$\sigma{_{T_c}}=0.32$\,K}.\label{Fig:CRaw}
\end{figure}

\begin{figure}
\center
\includegraphics[width=8cm]{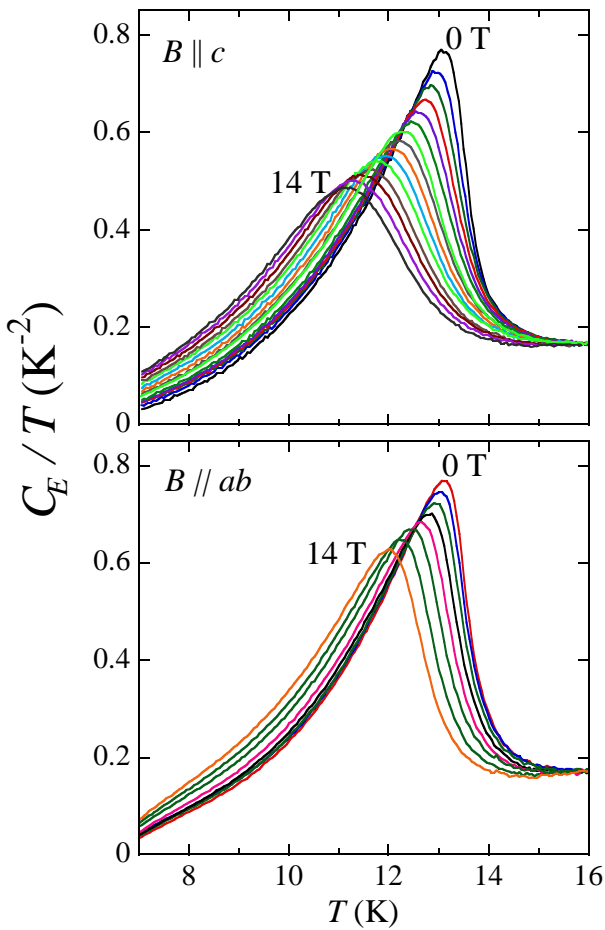}
\caption{(color online) Field dependence of the electronic heat capacity (i.e., raw data with the polynomial background
in Fig.\ \ref{Fig:CRaw} subtracted). In the upper panel the field is perpendicular to the planes and increases in 1\,T
intervals from 0 to 14\,T. In the lower panel the field is parallel to the planes and increases in 2\,T intervals from
0 to 14\,T.  Before subtraction of the background small ($<1$\% of the total) corrections were applied to normalize all
data to the same value at 16\,K.  This is necessary to correct for slight drift in the heating power.  Field sweeps at
this fixed temperature showed that $C(H)$ was constant (after correction for the thermocouple field dependence) within
resolution.} \label{Fig:C(H)_ab&c}
\end{figure}

\begin{figure}
\center
\includegraphics[width=8cm]{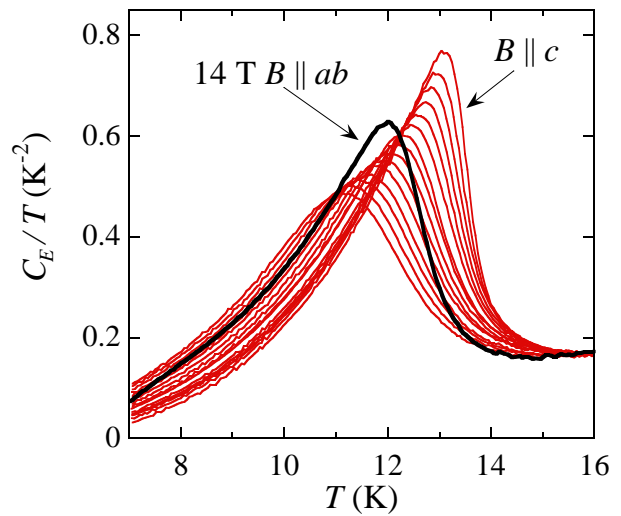}
\caption{(color online) Comparison of the field dependent heat capacity $C_E(H)$ for $B\|c$ with the 14\,T data for
$B\|ab$.} \label{Fig:C(H)_compare}
\end{figure}

The heat capacity of sample 1 (from the penetration depth study) is shown in Fig.\ \ref{Fig:CRaw}. In order to make the
evolution of the sample's electronic heat capacity with field clearer we have estimated the background (which includes
the samples phonon heat capacity and the addenda) in the following way.   We fitted the zero field data over the full
temperature range to a polynomial (5th order) to model the non-electronic terms plus the electronic heat capacity
calculated using a single gap alpha model. In the alpha model\cite{Padamsee73} strong-coupling effects are modelled by
multiplying the weak-coupling $s$-wave BCS temperature dependent energy gap $\Delta(T)$ by a constant $\alpha$, and
then the heat capacity is calculated in the standard way. The model works because the temperature dependent
strong-coupling corrections to $\Delta(T)$ are less important  than the zero temperature value of $\Delta(T)$ for the
behavior of $C$. For the purposes of the current study we can simply regard this as an entropy conserving construction
for estimating the field independent background. Note that the assumption of $s$ or $d$ wave behavior for $\Delta(T)$
makes insignificant differences in this range of temperature ($T/T_c\gtrsim 0.5)$ besides changing $\alpha$
\cite{TaylorCS07}.  The excellent fit of this model to the data is shown in Fig.\ \ref{Fig:CRaw}, gives
$\Delta_0/T_c=3.2$, which might indicate strong coupling (for elemental Pb $\Delta_0/T_c=2.25$).

In the subtracted data some broadening of the superconducting transition is evident, which comes from both sample
inhomogeneity (a distribution of $T_c$ values) and also strong thermal fluctuation effects (see later).  In order to
quantify the sample inhomogeneity effect in Fig.\ \ref{Fig:CRaw} we show a fit to the data with the alpha model
convoluted with a Gaussian $T_c$ distribution: i.e., $C(T)=\int_{-\infty}^{\infty} C(T,T_c) P(T_c)dT_c$ with the
probability distribution function $P(T_c)=\exp[-(T_c-T_c^0)^2/(2 \sigma^2_{T_c})]/\sqrt{2\pi\sigma^2_{T_c}}$ and
$C(T,T_c)$ fixed by the above non-convoluted alpha model fit.  As show in the inset to Fig.\ \ref{Fig:CRaw} this fits
the data well with $T_c^0 = 13.60$\,K and $\sigma{_{T_c}}=0.32$\,K. As fluctuation effect have not been included here
(for simplicity) this gives an upper limit on the sample inhomogeneity as $\sigma{_{T_c}}/T_c^0 \simeq 2$\%.  Note that
this distribution of $T_c$ in the sample has no intrinsic connection to the scattering possible responsible for the
$T^2$ behavior of the penetration depth.

The field dependence of the electronic heat capacity (i.e., raw data with the polynomial background in Fig.\
\ref{Fig:CRaw} subtracted) for the same sample measured in both field directions ($B\|c$ and $B\|ab$) is shown in Fig.\
\ref{Fig:C(H)_ab&c}. There is a clear striking difference in the behaviors in the two field directions.  For $B\|ab$
the behavior is close to that found for conventional type-II superconductors.  The superconducting anomaly shifts down
in temperature with the peak slightly decreasing in height (25\% in 14\,T), and the transition temperature width
remains almost constant. However, in the $B\|c$ direction the anomaly is very strongly broadened by the field with the
onset temperature remaining roughly field independent.  The anomaly height is reduced by $\sim$ 40\% in 14\,T.  This
qualitative difference is clearly illustrated in Fig.\ \ref{Fig:C(H)_compare} where we compare directly the $B\|c$ and
the 14\,T $B\|ab$ data. The reduction in anomaly height for 14\,T $B\|ab$ data is comparable to that found for 4\,T
$B\|c$ however, the broadening is not comparable at any field.

The behavior for $B\|c$ is similar to that found on high $T_c$ cuprate superconductors where the thermodynamics of the
superconducting transition are dominated by strong (critical) thermal fluctuation effects
\cite{InderheesSRG91,OverendHL94,JunodWTTRWM94,CarringtonMBCBMH97,SchillingFPWKC97}. Unlike conventional low-$T_c$
superconductors where the jump in $C$ at $T_c$ is almost a perfect example of a classic second order phase transition,
in YBa$_2$Cu$_3$O$_{6.9}$ ($T_c=93$\,K) the anomaly resembles the $\lambda$ anomaly found at the superfluid transition
of $^4$He which is well described by the 3D-XY critical fluctuation model \cite{LipaC83}.   The broadening (with little
or no reduction in onset temperature) and reduction of height of the anomaly is commonly observed in high-$T_c$
cuprates \cite{InderheesSRG91,OverendHL94,JunodWTTRWM94,CarringtonMBCBMH97,SchillingFPWKC97}. There is some controversy
regarding the best theoretical picture to describe this. Models based on finite size scaling within the critical 3D-XY
framework and fluctuation models based on the lowest Landau level (LLL) approximation have both been used
\cite{OverendHL94,JunodWTTRWM94,CarringtonMBCBMH97}. In principle, it should be expected that the finite size scaling
should work best at low fields (compared to $H_{c2}$ at $T=0$) and the LLL model at high fields \cite{OverendHL94}. The
finite size effects arise because the magnetic field introduces a new length scale $d_B =(\phi_0/\pi B)^{1/2}$, which
limits the growth of the physical size of the fluctuations as $T$ approaches $T_c$. When the transition is dominated by
such strong fluctuation effects the true phase transition at $H_{c2}$ is wiped out as the fluctuations in the order
parameter are much larger than the mean value over a large range of $(H,T)$ space close to the $H_{c2}(T)$ line
\cite{FisherFH91}. The heat capacity still displays a broad hump at the `$H_{c2}(T)$' line but the resistivity for
example does not become zero until lower temperature when the vortex lattice freezes and phase coherence is achieved.
The strong fluctuation effects in high $T_c$ cuprates are driven by their short coherence lengths, low dimensionality
and high critical temperatures \cite{FisherFH91}.  Compared to the cuprates FeSe$_{0.5}$Te$_{0.5}$ is relatively
isotropic and has a much lower $T_c$ but does have a comparably short coherence length.  The estimated irreversibility
field at $T=0$ of FeSe$_{0.5}$Te$_{0.5}$ is similar to that found for underdoped YBa$_2$Cu$_3$O$_{6.5}$ (Ref.
\onlinecite{Doiron-leyraudPLLBLBHT07}) or YBa$_2$Cu$_4$O$_8$ (Ref. \onlinecite{bangura08}).

The field broadening of $C(T)$ for $B\|c$ is in contrast to the more conventional behavior observed in 122 compounds
such as  BaFe$_2$(As$_{0.66}$P$_{0.33}$)$_2$ with $T_c$=30\,K \cite{Hashimoto0907.4399}, however evidence for strong
fluctuation effects have been seen in the field dependent heat capacity of the 1111 compound NdFeAsO$_{1-x}$F$_x$
(Ref.\ \onlinecite{Pribulova09}).  An unusual aspect of the behavior of FeSe$_{0.5}$Te$_{0.5}$ is that the fluctuations
effects appear to be strongly anisotropic.  In YBa$_2$Cu$_3$O$_{6.9}$ although the `$H_{c2}(T)$' and irreversibility
field $H_{\rm irr}$ are anisotropic, the behaviors in the two directions are identical apart from a difference in the
field scale (the heat capacity for 1\,T $\|c$ is identical to 8\,T$\|ab$) \cite{JunodWTTRWM94,SchillingFPWKC97}. Here
the behaviors are qualitatively different in the two field directions. Fluctuations effects are minimal for $B\|ab$ but
strong for $B\|c$. It would appear that in-plane field does not disrupt the phase coherence between the planes but for
$B\|c$ the effective dimensionality is strongly reduced and fluctuations effects enhanced.

\subsection{Resistivity}

\begin{figure}
\center
\includegraphics[width=7cm]{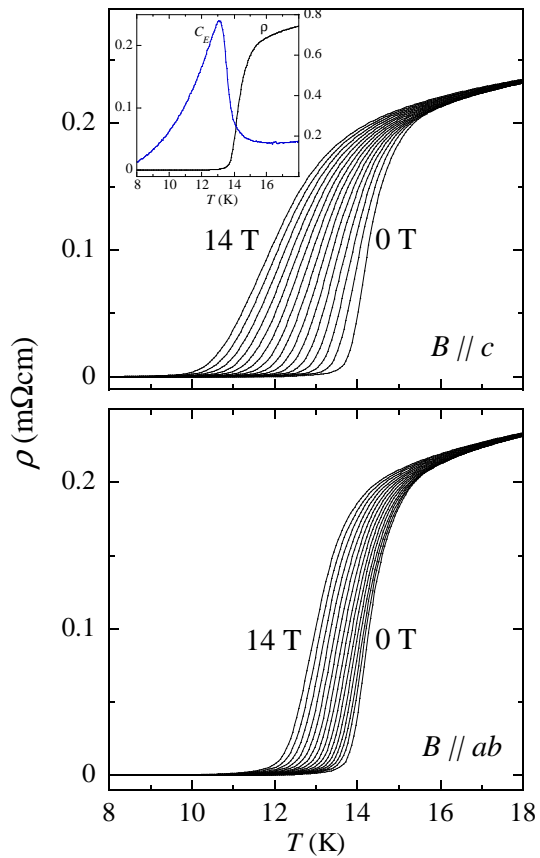}
\caption{(color online) Resistivity versus temperature for sample 1 for fields between 0 and 14\,T applied either
parallel or perpendicular to $c$.  The inset in the upper panel shows a direct comparison of the resistivity (left hand
scale)  and heat capacity (right hand scale) of the same sample (sample 1) in zero magnetic field.}
\label{Fig:Resistivity}
\end{figure}

The temperature and field dependence of the resistivity of sample 1 is shown in Fig.\ \ref{Fig:Resistivity}.  Although
there is again more broadening of the transition for $B\|c$ compared to $B\|ab$ it is much less evident than that seen
in the heat capacity data.  In the inset to Fig.\ \ref{Fig:Resistivity} we show a direct comparison between the zero
field heat capacity and resistivity data for the same sample. The mid point of the resistive transition is
$\sim$0.75\,K higher than the mid-point of the heat capacity.  The peak in the heat capacity is a further 0.5\,K lower
and corresponds to the point where the resistivity is $\sim$1\% of its value just above $T_c$. This difference is
understandable as a large reduction in the resistivity will be achieved when a sufficient amount of the sample to
create a percolation path becomes superconducting. Often this amount will be much less than 50\% but depends on the
spatial distribution of $T_c$ in the sample.

\subsection{Torque}

\begin{figure}
\center
\includegraphics[width=8cm]{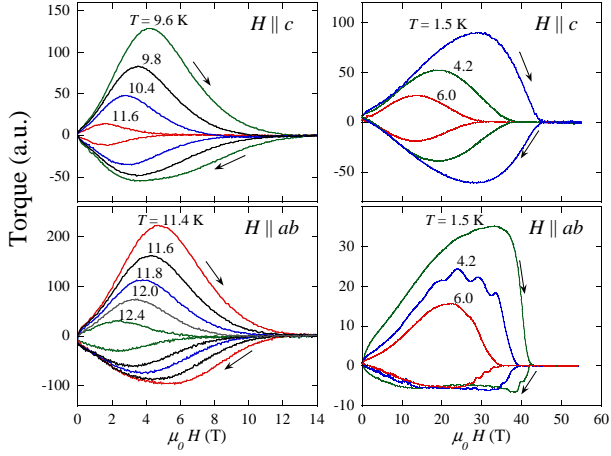}
\caption{(color online) Torque hysteresis loops.  The left two panels show data in quasi-dc field for a small piece cut
from sample 1.  The right two panels show data taken in pulsed field for a second sample. The field directions are
actually $\sim 5^\circ$ from the symmetry directions indicated on the figure.} \label{Fig:torqueLoops}
\end{figure}

Torque data in quasi-static fields up to 14\,T for a small piece cut from the heat capacity sample is shown in Fig.\
\ref{Fig:torqueLoops} at angles of approximately $5^\circ$ from $B\|c$ and $B\|ab$. The angle offset is necessary
because at the symmetry points the torque is exactly zero.  These small offset make negligible difference to the
derived irreversibility fields as the anisotropy is not strong.  Data taken in pulsed fields for a different sample
from the same batch is shown in Fig.\ \ref{Fig:torqueLoops}.  In both cases the torque is strongly hysteretic
indicating strong pinning. In the pulsed field $B\|c$ data oscillations are evident which likely arise from flux jumps.
In all data, the torque loops close at within our resolution at an irreversibility field ($H_{\rm irr}$), and no
further kinks at higher field (which might mark $H_{c2}$) are evident.

\subsection{Phase diagram}

\begin{figure}
\center
\includegraphics[width=8cm]{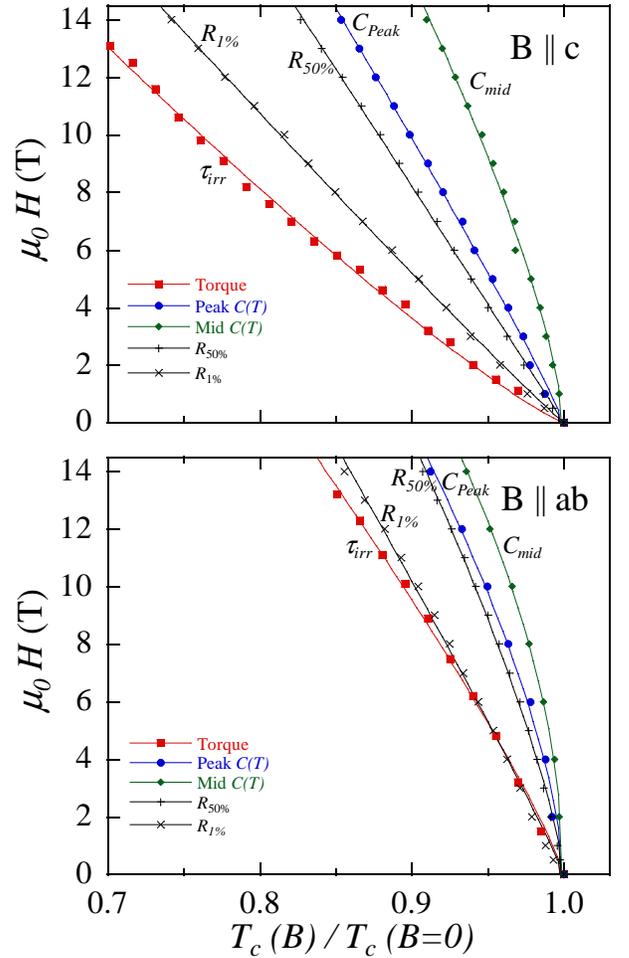}
\caption{(color online) $H-T$ phase diagram for FeSe$_{0.5}$Te$_{0.5}$ close to $T_c$ derived from heat capacity,
resistivity and torque measurements. For the heat capacity the temperature of the peak in $C(T)$ and the mid point of
the rise above the peak the transition are plotted.  For the resistivity, $R_{50\%}$ and $R_{1\%}$ are the points at
which the resistivity has fallen to 50\% and 1\% of extrapolated normal state resistivity above $T_c$ respectively. For
the torque the point where the hysteresis loop closes (within our resolution), i.e, the irreversibility field
$\mu_0H_{irr}$ is plotted. The lines are fits to the power-law $H=H_0(1-T/T_c)^n$, to guide the eye. }
\label{Fig:phasediagramnearTc}
\end{figure}

A compilation of the data derived from the heat capacity, resistivity and torque measurements is plotted in Fig.\
\ref{Fig:phasediagramnearTc} as a function of reduced temperature $T_c(B)/T_c(B=0)$.  It is important to realize that
in general these different experimental techniques will measure different points in the $(H,T)$ phase diagram.   For a
superconductor without strong fluctuation effects there will be a step in the heat capacity at $T_c$ which will be
broadened by inhomogeneity and will shift down in temperature under applied field marking the $H_{c2}(T)$ line.  The
width of the broadened step in field will be related to the width in zero field and the slope of $H_{c2}(T)$, i.e.,
$\Delta T_c(H) = H_{c2} \Delta T_c(0)/(dH_{c2}/dT)$.  In this case the mid point of the rise of the $C(T)$ can be used
to determine $H_{c2}(T)$.  However, in the present case, as the broadening of $C(T)$ in field cannot be explained by
inhomogeneity alone there must also be strong thermal fluctuation effects as described above. In the case of phase
transitions where fluctuation effects are strong, the peak in $C(T)$ is usually used to define the phase transition
point, and so for the present case this likely gives the best estimate of $H_{c2}(T)$.   At the point at which the
magnetization becomes irreversible (the irreversibility line) the sample will acquire a finite critical current.  Hence
the irreversibility line as measured by our torque measurements marks the point where the critical current takes a
finite value - the size of which depends on the resolution of the torque measurements.  It should be expected that this
irreversibility line is close to the point where the resistivity becomes zero as this is also where the critical
current becomes equal to the resistivity measurement current.  In the case that the irreversibility line and $H_{c2}$
are not in close proximity it is difficult to know what point in the resistivity curve corresponds to $H_{c2}(T)$.  For
example, if the vortex viscosity is very low the resistivity in the mixed state will be close to the normal state value
and there will be almost no anomaly at $H_{c2}(T)$ \cite{CarringtonMT96,GeshkenbeinIM98}.

In the present case, the peak in $C(T)$ coincides reasonably well with the point on the resistivity curve where the
resistance has fallen to 50\% of its extrapolated normal state value ($R_{50\%}$). This is the case for \textit{both}
field directions (see Fig.\ \ref{Fig:phasediagramnearTc}). Hence is it reasonable to conclude that both of these two
quantities are measuring the temperature dependence of $H_{c2}$.  The 1\% resistivity point ($R_{1\%}$) is better
correlated with the torque irreversibility field as expected from the above discussion.

For $B\|c$ there is clearly a very marked difference between the irreversibility line derived from the torque
measurements and the thermodynamic transition measured by specific heat. The average slope of the torque
irreversibility field $dH_{irr}^{\|c}/dT = -44$\,T/K whereas for the peak in $C(T)$ it is $-95$\,T/K and the midpoint
$-150$\,T/K. Hence estimates of $H_{c2}$ from the irreversibility field for this field direction will serious
underestimate the true values. Although the data are better fitted by a power law $H = (1-T/T_c)^n$ as indicated in the
figure rather than a linear relation, the average slope still provides a useful rough comparison. The large separation
between the irreversibility field and the peak in the specific heat which we expect to be close to the mean-field
$H_{c2}$ line is again reminiscent of the behavior found in the high-$T_c$ cuprates and is strongly suggestive of
strong fluctuation effects \cite{InderheesSRG91,OverendHL94,JunodWTTRWM94,CarringtonMBCBMH97,SchillingFPWKC97}.   For
$B\|ab$ the lines are much closer together. The corresponding values for the average slopes are $-90$\,T/K, $-160$\,T/K
and $\-200$\,T/K respectively.

The anisotropy of $H_{c2}$ derived from the specific heat measurements is considerably less than that derived from the
torque irreversibility field.  Taking the peak in $C(T)$ as the most reliable measure of $H_{c2}(T)$ we see that for
$B\|c$ it is almost linear with temperature whereas for $B\|ab$ there is some curvature close to $T_c(0)$. Above $\sim
8$\,T where $H_{c2}^{\|ab}(T)$ is linear the anisotropy is small, $\frac{dH_{c2}^{\|ab}}{dT}/\frac{dH_{c2}^{\|c}}{dT} =
1.2$.

\begin{figure}
\center
\includegraphics[width=8cm]{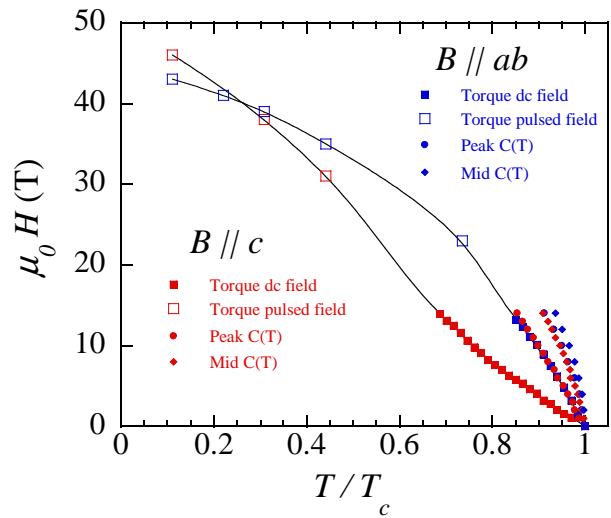}
\caption{(color online) $H-T$ phase diagram for FeSe$_{0.5}$Te$_{0.5}$ over the full range.  The torque data mark the
irreversibly line whereas the heat capacity data $C(T)$ are two different estimates of the behavior of $H_{c2}$. The
lines are guides to the eye.} \label{Fig:Highfieldphasediagram}
\end{figure}

Braithwaite \emph{et al.} \cite{BraithwaiteLKS10} have also reported a difference between critical fields determined by
resistivity or heat capacity measurements for FeSe$_{0.48}$Te$_{0.52}$. In their case, they found $R_{50\%}$ and the
mid-point of the $C(T)$ curve approximately coincided for $H\|ab$ but not for $H\|c$. Their heat capacity data showed
almost no anisotropy. $C(T)$ data for $\mu_0H$=9\,T was almost independent of the field direction. It is likely that
the differences in the $H\|ab$ data compared to ours is due to the much wider transition width of their sample. The
zero field $C(T)$ data in Ref.\ \onlinecite{BraithwaiteLKS10} are approximately three times wider than those here. The
field broadened transition width of our data for  $\mu_0H$=9\,T $\|c$ is comparable to the zero field width in Ref.\
\onlinecite{BraithwaiteLKS10}.

The phase diagram over a wide temperature-field range is shown in Fig.\ \ref{Fig:Highfieldphasediagram}.  As reported
previously \cite{BraithwaiteLKS10,Khim10} the irreversibility fields for the two field directions cross at low
temperature with the anisotropy approaching unity. This and the general shape of these $H_{\rm irr}(T)$ curves, has
been interpreted as evidence that $H_{c2}$ is Pauli limited \cite{BraithwaiteLKS10,Khim10}.  However, given the
difference between $H_{\rm irr}$ and $H_{c2}$ revealed by the lower field heat capacity data, it is possible that this
high field behavior is also strongly influenced by thermal fluctuation effects and vortex pinning.  High field heat
capacity measurements will be necessary to determine whether the temperature dependence and anisotropy of $H_{c2}$
follow that of $H_{\rm irr}$ at high field and low temperature.

\section{Pair-breaking}
As mentioned in the introduction, Kogan \cite{Kogan10} has calculated that in anisotropic superconductors with strong
pair-breaking there should be a universal relationship between the co-efficient of the $T^2$ term in the low
temperature penetration depth, the height of the mean-field part of the specific heat jump $\Delta C_{\rm MF}$, and the
slope of the $H_{c2}$ near $T_c$. In SI units the relation is
\begin{equation}
\Omega=\frac{\Delta C_{\rm MF} A^2 T_c^4}{\left|\frac{dH_{c2}}{dT}\right|_{T_c}} = \frac{\phi_0}{4\pi} \label{Eq:Kogan}
\end{equation}
Note here that $\Delta C_{\rm MF}$ is units of J m$^{-3}$ K$^{-1}$ and $\phi_0$ is the flux quantum. Fixing $n=2$ in
Eq.\ \ref{Eq:lamTfit} we get $A = 10.3$\,\AA/K for the $T^2$ slope of the low temperature $\lambda(T)$ data.  As argued
above, the peak in $C(T,B)$ is likely to provide the best estimate of $H_{c2}(T)$ and from this we get
$\mu_0\frac{dH_{c2}}{dT}=6.9$\,T/K for $B\|c$ and $T$ between $0.85$ and $0.97\,T_c$.  As our specific heat data is not
in absolute units we estimate $\Delta C_{\rm MF}$ from the data of Tsurkan \emph{et al.}\cite{Tsurkan1006.4453} for the
specific heat of FeSe$_{0.5}$Te$_{0.5}$. This data is very similar to our own with regard to sharpness and value of
$T_c$ and slope of $H_{c2}$, and from this we estimate $\Delta C_{\rm MF}=0.42$\,Jmol$^{-1}$K$^{-1}$.  Taking the mid
point of the heat capacity transition $T=13.6$\,K as $T_c$, we get from Eq.\ \ref{Eq:Kogan}, $4\pi\Omega/\phi_0=0.65
\pm 0.15$ which is indeed close to unity as predicted by the Kogan theory (the uncertainty is dominated by that in
$T_c$).  Hence, it is likely that the $T^2$ behavior of the penetration depth does result from strong pair breaking.
However, this does not rule out the possibility of intrinsic nodes if the material could be made sufficiently clean.

\section{Conclusions}

We have presented results for the temperature dependence of the magnetic penetration depth, heat capacity and magnetic
torque for highly homogeneous single crystal samples of  Fe$_{1.0}$Se$_{0.44(4)}$Te$_{0.56(4)}$.  The penetration depth
data display a power law behavior $\Delta\lambda\propto T^n$ with $n=2.2\pm0.1$, which is similar to some 122
iron-arsenides. It is likely this results from the influence of strong pair breaking with a sign-changing pairing state
which may or may not have intrinsic nodes \cite{vorontsov:140507,Bang09,Kogan10,Gordon10}. Heat capacity measurements
have shown clear evidence for the presence of strong thermal fluctuations, at least for relatively low fields
($\mu_0H\leq 14$\,T).  The behavior is reminiscent of the high $T_c$ cuprates, and like in these compounds we show that
there is a sizable vortex liquid regime between the irreversibility line and the upper critical field. Unusually, the
fluctuations influence the behavior with field along $c$ much greater than for field along the planes.

\section*{Acknowledgements}
We thank Dr.\ John Bacsa for single crystal XRD studies and Mr.\ Chris Ireland for the EDX measurements.  We also thank
Vladimir Kogan for drawing our attention to his pair-breaking theory.  We acknowledge financial support from the Royal
Society and EPSRC. Part of this work has been done with the financial support of EuroMagNET II under the EU contract
no.228043.
\bibliography{FeTeSe}
\bibliographystyle{apsrevac}
\end{document}